\documentclass[12pt]{article}
\usepackage{epsfig}
\usepackage{comment}
\usepackage{latexsym}
\usepackage{color}
\newcommand{\mysquare}[0]{\raise-.2ex\hbox{{\Large$\Box$}}}
\def\lsim{\mathrel{\rlap {\raise.5ex\hbox{$ < $}}
{\lower.5ex\hbox{$\sim$}}}}
\def\gsim{\mathrel{\rlap {\raise.5ex\hbox{$ > $}}
{\lower.5ex\hbox{$\sim$}}}} \topmargin -1.5cm \textheight=22.5cm \textwidth=16.5cm
\catcode`\@=11
\newcount\hour
\newcount\minute
\newtoks\amorpm
\hour=\time\divide\hour by60 \minute=\time{\multiply\hour by60 \global\advance\minute by-\hour}
\edef\standardtime{{\ifnum\hour<12 \global\amorpm={am}%
        \else\global\amorpm={pm}\advance\hour by-12 \fi
        \ifnum\hour=0 \hour=12 \fi
        \number\hour:\ifnum\minute<10 0\fi\number\minute\the\amorpm}}
\edef\militarytime{\number\hour:\ifnum\minute<10 0\fi\number\minute}
\def\draftlabel#1{{\@bsphack\if@filesw {\let\thepage\relax
   \xdef\@gtempa{\write\@auxout{\string
      \newlabel{#1}{{\@currentlabel}{\thepage}}}}}\@gtempa
   \if@nobreak \ifvmode\nobreak\fi\fi\fi\@esphack}
        \gdef\@eqnlabel{#1}}
\def\@eqnlabel{}
\def\@vacuum{}
\def\draftmarginnote#1{\marginpar{\raggedright\scriptsize\tt#1}}
\def\draft{\oddsidemargin -.2truein
        \def\@oddfoot{\sl preliminary draft \hfil
        \rm\thepage\hfil\sl\today\quad\militarytime}
        \let\@evenfoot\@oddfoot \overfullrule 3pt
        \let\label=\draftlabel
        \let\marginnote=\draftmarginnote
   \def\@eqnnum{(\theequation)\rlap{\k

 ern\marginparsep\tt\@eqnlabel}%
\global\let\@eqnlabel\@vacuum}  }

\newcommand{\be}[0]{\begin{equation}}
\newcommand{\ee}[0]{\end{equation}}
\newcommand{\ba}[0]{\begin{eqnarray}}
\newcommand{\ea}[0]{\end{eqnarray}}

\def\bs{\begin{subequations}}
\def\es{\end{subequations}}

\def\thebibliography#1{%
\vskip 0.5cm \centerline{\bf \Large References}
\list{%
[\arabic{enumi}]}{\settowidth\labelwidth{[#1]} \leftmargin\labelwidth \advance\leftmargin\labelsep
\usecounter{enumi}}
\def\newblock{\hskip .11em plus .33em minus .07em}
\sloppy\clubpenalty4000\widowpenalty4000 \sfcode`\.=1000\relax}

\renewcommand{\theequation}{\arabic{section}.\arabic{equation}}

\renewcommand{\section}{\setcounter{equation}{0}\@startsection
{section}{1}{0mm}{-\baselineskip}{0.5\baselineskip} {\normalfont\Large\bfseries}}

\renewcommand{\subsection}{\@startsection
{subsection}{2}{0mm}{-\baselineskip}{0.5\baselineskip} {\normalfont\large\bfseries}}

\renewcommand{\subsubsection}{\@startsection
{subsubsection}{3}{0mm}{-\baselineskip}{0.5\baselineskip} {\normalfont\normalsize\slshape}}

\usepackage{amsmath}
\usepackage{amssymb,amsfonts}
\usepackage{graphicx}
\usepackage{cite}

\renewcommand{\and}{\mbox{and}}

\begin{document}
\begin{titlepage}
\begin{flushright}
LPTENS--10/02,
January 2010
\end{flushright}


\begin{centering}
{\bf\huge String Models with Massive boson-fermion Degeneracy}\\

\vspace{4mm}
 {\Large Ioannis~G.~Florakis \\
 }

\vspace{2mm}

$^1$ Laboratoire de Physique Th\'eorique,
Ecole Normale Sup\'erieure,$^\dagger$ \\
24 rue Lhomond, F--75231 Paris cedex 05, France\\
\vspace{2mm}
{\em  Ioannis.Florakis@lpt.ens.fr}

\vskip .1cm

 \vspace{3mm}

{\bf\Large Abstract}

\end{centering}
\begin{quote}

We discuss constructions of string vacua with a novel Massive Spectral boson-fermion Degeneracy Symmetry (\emph{MSDS}) as candidate vacua able to describe the early non-geometrical era of the universe.
\begin{center}
\emph{ Based on a talk given at the Workshop on Cosmology \& Strings,\\ Corfu Institute, Greece, Sept 12, 2009}.
\end{center}

\noindent

\end{quote}
\vspace{5pt} \vfill \hrule width 6.7cm \vskip.1mm{\small \small \small \noindent $^\ast$
$^\dagger$\ Unit{\'e} mixte  du CNRS et de l'Ecole Normale Sup{\'e}rieure associ\'ee \`a
l'Universit\'e Pierre et Marie Curie (Paris
6), UMR 8549.}

\end{titlepage}
\newpage
\setcounter{footnote}{0}
\renewcommand{\thefootnote}{\arabic{footnote}}
 \setlength{\baselineskip}{.7cm} \setlength{\parskip}{.2cm}

\setcounter{section}{0}

\section{String Cosmology and Initial Vacuum Selection}

It is well-known from classical cosmology that, extrapolating general relativity backwards in time, one encounters a time-like singularity. Physically, such gravitational singularities arise as artifacts of our extrapolation of classical physics into a region where quantum gravitational effects become dominant. One would expect that, by replacing general relativity with a more fundamental theory of quantum gravity -such as string theory-, will eventually smear out these pathologies. This should be compared to the well-known example from quantum mechanics where quantum fluctuations protect a particle from falling into a singular potential.

Studying string theory in the strong curvature and high temperature regime provides a fundamentaly different picture of spacetime than the one expected from the field theory approximation. Classical notions, such as geometry, topology and even the dimensionality of space are only well-defined notions in the effective low-energy approximation of string theory \cite{CosmoTopologyChange} (also \cite{MSDS}, \cite{reducedMSDS} and references therein). In the strong curvature and high temperature phase string theory, where field-theoretic notions break down, new purely stringy phenomena take place. Essentially, this is because of the presence of extended symmetry points in the moduli space of string theory. A thorough understanding of string cosmology \cite{BV} in these regimes may be the first step in dynamically connecting the early stringy phase of the universe with standard cosmological and phenomenological theories at late times.

This ambitious program is highly dependent on the particular choice of the Initial String Vacuum. Candidate vacua should be formulable as exactly solvable CFTs admitting a thermal interpretation, with the compactification radii (temperature) close to the string scale (Hagedorn point). Ideally, good candidates would be free of tachyonic/Hagedorn divergencies to enable a perturbative treatment of the backreaction. Finally, the initial vacuum should be continuously connected to 4-dimensional semi-realistic vacua with $\mathcal{N}=1$ spontaneously broken supersymmetry at late cosmological times and with chiral matter in a semi-realistic GUT gauge group, such as $SO(10)$.

The presence of tachyonic (or Hagedorn) instabilities \cite{Hag} signals a stringy phase transition into a new vacuum (thermodynamical phase) of non-trivial winding number. Therefore, we could try to directly construct vacua with non-trivial winding number that are free of the above pathologies. Since the absence of physical tachyons from the string spectrum of $2d$-vacua is essentially equivalent (see \cite{KutSeiberg}, \cite{Misaligned}) to an asymptotic degeneracy $n_B-n_F\rightarrow 0$ in the difference of the numbers of bosonic minus fermionic states for large mass levels, the desired initial vacua should be such that supersymmetry is broken in the low-energy spectrum, whereas at large mass levels the boson-fermion degeneracy is asymptotically restored. These considerations provided the motivation for the discovery and construction of vacua \cite{MSDS} with a novel \emph{Massive Spectral boson-fermion Degeneracy Symmetry} (MSDS). This family of string vacua satisfy several of the above requirements, and will be introduced below.

\section{Maximally Symmetric MSDS Vacua}

In their non-Euclidean version, the MSDS vacua correspond to 1+1-dimensional compactifications $\mathcal{M}^{2}\times K$ on a compact space $K$, that is described by a $\hat{c}=8$ CFT of free fermions, where all 8 of the internal coordinates are compactified at the ``fermionic" radius $R_i=\sqrt{\alpha'/2}$, where the chiral (compact) worldsheet bosons can be consistently fermionized. The left-moving worldsheet degrees of freedom (e.g. for Type II theories) in the RNS formalism are the 2 lightcone supercoordinates $(\partial X^{0,L}, \Psi^{0,L})$, the ghost and superghost systems $(b,c), (\beta,\gamma)$ and the 8 transverse supercoordinates $(\partial X^I, \chi^I)$, where $I=1,\ldots ,8$. The right-moving side has similar worldsheet d.o.f. for Type II theories, whereas in the Heterotic case, we add 16 additional (complex) free fermions $\psi^A$, with $A=1,\ldots,16$, to cancel the right-moving conformal anomaly. Fermionization then amounts to realizing the abelian $U(1)^8$ current algebra of transverse bosons $\partial X^I$ in terms of free worldsheet fermions $i\partial X^I(z) =y^I \omega^I(z)$. The fermions $\{\chi^I,y^I,\omega^I\}$ then realize a global $G=\widehat{SO}(24)_{k=1}$ affine algebra. Fermionization of the right-movers proceeds similarly. However, there are extra constraints arising from worldsheet supersymmetry which is now realized non-linearly among the free fermions $\delta\psi^a\sim f_{abc}\psi^b\psi^c$, with $a=1,\ldots, 24$. The condition that the latter is a \emph{real} supersymmetry imposes that the original symmetry $G$ is gauged down to a subgroup $H$, such that $G/H$ is a symmetric space and, therefore, forces the free fermions to transform in the adjoint representation of the semi-simple Lie algebra of $H$ with $\dim{H}=24$. When these conditions are met the local currents $J_a=f_{abc}\psi^b\psi^c$ together with the 2d energy-momentum tensor $T_B$ and  supercurrent $T_F$ close into an $\mathcal{N}=1$ worldsheet superconformal algebra. Several gaugings are possible $SU(2)^8$, $SU(5)$, $SO(7)\times SU(2)$, $G_2\times Sp(4)$, $SU(4)\times SU(2)^3$, $SU(3)^3$. In these constructions we only use the gauging of maximal rank $H=SU(2)^8_{k=2}$.

The prototype MSDS vacua \cite{MSDS} are obtained by assigning the same spin-structures $(a,b)$ to all $24$ (real) free worldsheet fermions. The modular invariant partition functions for Type II theories is then:
\begin{align}
	Z_{\textrm{II}} =\frac{1}{2^2}\sum\limits_{a,b=0,1}{(-)^{a+b}\frac{\theta[^a_b]^{12}}{\eta^{12}}}\sum\limits_{\bar{a},\bar{b}=0,1}{(-)^{\bar{a}+\bar{b}}\frac{\bar\theta[^{\bar{a}}_{\bar{b}}]^{12}}{\bar\eta^{12}}} =(V_{24}-S_{24})(\bar{V}_{24}-\bar{S}_{24})=576,
\end{align}
and for Heterotic theories:
\begin{align}
	Z_{\textrm{het}} =\frac{1}{2}\sum\limits_{a,b=0,1}{(-)^{a+b}\frac{\theta[^a_b]^{12}}{\eta^{12}}}~\Gamma[H_R] =24\times\left(d[H_R]+[\bar{j}(\bar{z})-744]\right).
\end{align}
The number of anti-chirally massless states $d[H_R]$ is equal to $1128$ for $H_R=SO(48)$ and $744$ for $H_R=SO(32)\times E_8$ or $H_R=E_8^3$. The combination $(j(z)-744)$ involving the Klein invariant function is eliminated after integration (or by imposing level-matching). The astonishing property of MSDS vacua is that there is an equal number of bosons and fermions $n_B=n_F$ at all massive states, with the exception of the massless states which, in these maximal models where the $H_L\times H_R$ gauge symmetry is unbroken, are all bosonic. The MSDS structure of the partition function in general models can be generally shown to be of the form $Z=p+q(\bar{j}-744)$, where $p,q\in\mathbb{Z}$ and $p=n_B-n_F$ at the massless level.
These maximal models satisfy in an \emph{exact} way the required boson-fermion degeneracy, and one may be tempted to call this structure ``massive supersymmetry". In fact, we will argue below that MSDS arises as a purely stringy ``gauging" of supersymmetry.

The MSDS structure arises from the generalized Jacobi identity $V_{24}-S_{24}=24$, that replaces the ``abstrusa" identity $V_8-S_8=0$ expressing the triality of $SO(8)$ in the case of conventional supersymmetry. The chiral nature of these identities hints at the presence of a \emph{chiral} operator $j^{\textrm{MSDS}}(z)$ responsible for the mapping of \emph{massive bosonic to massive fermionic} representations, while \emph{leaving massless states invariant}. This operator is found to be:
\begin{align}
	j_{\alpha}(z) = e^{\frac{1}{2}\Phi-\frac{i}{2}H_0}~C_{24,\alpha}(z).
\end{align}
where $C_{24,\alpha}$ is the spin-field of $SO(24)$ with positive chirality. 
The ghost dressing contributes $-5/8+1/8$ to the conformal weight and $j_{\alpha}$ acts as a weight $(1,0)$-current. Its zero mode $Q_{\textrm{MSDS}}=\oint{\frac{dz}{2\pi i}~j_{\alpha}(z)}$ defines a conserved MSDS charge ensuring the mapping of states at the same mass level. Indeed, considering the primary operators generating the vectorial $\textbf{V}=e^{-\Phi}\hat{\psi}$ and the spinorial $\textbf{S}=e^{-\frac{1}{2}\Phi-\frac{i}{2}H_0}S_{24,\alpha}$ representations and calculating their OPEs with the MSDS current we find that the massless states in the vectorial do not transform $j(z)\textbf{V}(w)=\textrm{regular}$, whereas the first massive descendants of the vectorial family generated by $[\textbf{V}]_{(1)}=e^{-\Phi}\mathcal{D}\hat\psi$ transform into the spinorial $\textbf{S}$ and vice-versa. Here $\hat{\psi}\equiv \psi^a\gamma^a$ denotes contraction with the $\gamma$-matrices of $SO(24)$ and $\mathcal{D}\equiv\partial+\hat{J}$ is a covariant-like derivate associated to the gauging of supersymmetry and $\hat{J}\equiv\hat\psi\hat\psi$. As a result, only the 24 (chiral) massless states $\psi^a$ are unpaired and contribute to the character difference $V_{24}-S_{24}=24$.

The chiral $\mathcal{N}=1$ superconformal algebra of $\{T_B,T_F,J_a\}$ admits a spectral flow generated by the spinorial MSDS currents $j_{\alpha}(z)$. One may calculate the non-trivial OPEs between the currents and show they are expressed in terms of the covariant derivatives $\mathcal{D}$:
\begin{align}
	& j_{\alpha}(z)j_{\beta}(w)\sim e^{\Phi}\left(\frac{\Psi^{-}(w)}{(z-w)^3}+\frac{\mathcal{D}\Psi^{-}(w)}{(z-w)^2}+\frac{\frac{1}{2}\left(\mathcal{D}^2+(\mathcal{D}\hat{J})\right)\Psi^{-}(w)}{z-w}\right)_{\alpha\beta}\\
	& j_{\alpha}(z)\tilde{j}_{\dot{\beta}}(w)\sim e^{\Phi}\left(\frac{\hat\psi(w)}{(z-w)^3}+\frac{\mathcal{D}\hat\psi(w)}{(z-w)^2}+\frac{\frac{1}{2}\left(\mathcal{D}^2+(\mathcal{D}\hat{J})\right)\hat{\psi}(w)}{z-w}\right)_{\alpha\dot{\beta}},
\end{align}
where $\tilde{j}_{\dot{\beta}}=e^{\frac{1}{2}\Phi+\frac{i}{2}H_0}S_{\dot{\beta}}$ and $\Psi^{-}\equiv \frac{1}{\sqrt{2}i}(\Psi^{0}-\Psi^{L})$.  Comparing the MSDS current $j_{\alpha}$ with the spacetime supercurrent in the $\left(+\frac{1}{2}\right)$-picture $q_{\hat\alpha}=e^{\frac{1}{2}\Phi-\frac{i}{2}H_0}C_{8,\hat\alpha}\partial X^I+\ldots$, with $\hat\alpha$ being the spinorial index of the $SO(8)$ holonomy group, we see that MSDS can be regarded as an ``enhanced supersymmetry" at the fermionic point of moduli space. From this point of view, MSDS symmetry is a purely stringy gauging of supersymmetry. 

The MSDS models also admit a very natural thermal interpretation by identifying the modulus coupling to the total fermion number with the temperature of the canonical ensemble. A second modulus couples to the right-moving fermion number and is interpreted as a supersymmetry breaking scale. The geometrical fluxes associated to the Euclidean time direction are then shown to correspond to imaginary chemical potentials for the thermodynamical system.

To have a sensible phenomenology, we must ensure that MSDS vacua are eventually connected to semirealistic $\mathcal{N}=1$ chiral models at late cosmological times. For this purpose, we consider orbifolds that preserve the MSDS structure. A complete classification of MSDS-preserving $\mathbb{Z}_2^N$-orbifold constructions was given in \cite{reducedMSDS}. The result is that whenever the boundary conditions and GSO-projections respect the \emph{global, chiral} definition of the spectral-flow operator $j_{\alpha}$ and if, in addition to other consistency conditions, the boundary condition vectors $b_a$  (defined from $\psi^a\rightarrow -e^{i\pi b_a}\psi^a$) satisfy the extra \emph{holomorphic} constraint $n_L(b)=0(\textrm{mod}~8)$  the resulting string vacua preserve the MSDS structure. A similar construction is also possible for more general $\mathbb{Z}_k$-orbifolds.

The above constructions permit the complete classification of Type II and Heterotic MSDS vacua. Remarkably, the restrictions arising from imposing MSDS structure are enormously constraining. In particular, semirealistic Heterotic MSDS models based on $SO(10)$ GUT gauge group have been constructed in 2 dimensions with spontaneously broken $\mathcal{N}=1$ supersymmetry in the presence of specific gravito/magnetic fluxes \cite{reducedMSDS},\cite{SO10MSDS}. As an example, consider the following $T^8/\mathbb{Z}_2\times\mathbb{Z}_2$ model \cite{SO10MSDS}:
\begin{align}\nonumber
	Z=&\frac{1}{\eta^{12}\bar\eta^{24}}\frac{1}{2^8}\sum\limits_{\Gamma_\alpha, \Delta_\beta}\sum\limits_{h_a, g_b}{(-)^{a+b+HG+{\Phi}}\,\theta[^a_b]\theta[^{a+h_1}_{b+g_1}]\theta[^{a+h_2}_{b+g_2}]\theta[^{a-h_1-h_2}_{b-g_1-g_2}]}\\  
&\times{\Gamma_{(8,8)}\left[^{a,k}_{b,l}\right|\left.^{P,h_1,h_2,\psi}_{Q,g_1,g_2,\omega}\right]}~\underbrace{\bar\theta[^k_l]^5}_{SO(10)}\underbrace{\bar\theta[^{k+h_1}_{l+g_1}]\bar\theta[^{k+h_2}_{l+g_2}]\bar\theta[^{k-h_1-h_2}_{l-g_1-g_2}]}_{U(1)\times U(1)\times U(1)}\times
	\underbrace{\bar\theta[^\rho_\sigma]^4\bar\theta[^{\rho+H}_{\sigma+G}]^4}_{SO(8)\times SO(8)},
\end{align}
where $\Gamma_\alpha=(a,k,\rho),~ \Delta_\beta=(b,l,\sigma),~ h_a=(P,h_1,h_2,\psi,H),~ g_b=(Q,g_1,g_2,\omega,G)$. The lattice is also expressed in terms of $\vartheta$-functions:
\begin{align}
{\Gamma_{(8,8)}}  {\equiv \theta[^{a+P}_{b+Q}]^5\theta[^{a+P+h_1}_{b+Q+g_1}]\theta[^{a+P+h_2}_{b+Q+g_2}]\theta[^{a+P-h_1-h_2}_{b+Q-g_1-g_2}]} 
{\times\,\bar\theta[^{k+\psi}_{l+\omega}]^5\bar\theta[^{k+\psi+h_1}_{l+\omega+g_1}]\bar\theta[^{k+\psi+h_2}_{l+\omega+g_2}]\bar\theta[^{k+\psi-h_1-h_2}_{l+\omega-g_1-g_2}]}.
\end{align}
There are $2^{29}$ choices for the modular invariant phase $\Phi$, corresponding to possible GSO projections. For some choices, the model is naively supersymmetric, whereas it is possible to choose the GSOs so that supersymmetry is (spontaneously) broken, while preserving the MSDS structure. The MSDS current invariant under the $\mathbb{Z}_2$'s can be taken to be $j(z)=e^{\frac{1}{2}\Phi-\frac{i}{2}H_0} C_2 C_2 C_2 C_2 C_2 C_2 C_2 C_{10}$.

An analysis of the moduli space of marginal deformations of the current-current form $J_I(z)\times\bar{J}_J(\bar{z})$, shows that these MSDS vacua are \emph{continuously} connected \cite{DeformedMSDS} to semirealistic four-dimensional $\mathcal{N}=1$ superstring vacua with a gauge group containing a semi-realistic $SO(10)$-GUT factor \cite{SO10MSDS}.

Studying MSDS models and their marginal deformations is only the first step in the ambitious project of connecting string cosmology with late-time phenomenology. Assuming that, in its early phase, the universe has at least $\hat{c}=8$ dimensions compactified close to the string scale, it seems very tempting to consider MSDS vacua as natural candidates for the description of the early non-geometric era of the universe. What is more, MSDS models are the first examples of vacua where a purely stringy enhancement of supersymmetry, in terms of a chiral algebra, entirely determines the spontaneous supersymmetry breaking. Their most attractive feature is their continuous connection with semi-realistic superstring models in four dimensions.

\section{Acknowledgements}

	I would like to aknowledge fruitful discussions with C.~Kounnas, J.~Rizos, N.~Toumbas and J.~Troost. I am also grateful to K.~Dienes for his comments on boson-fermion degeneracy in $2d$-vacua. Finally, it is a pleasure to thank the organizers and especially Ms. I.~Moraiti for her assistance with the administrative and logistic issues that arose.


\end{document}